\documentclass{article}

\usepackage{microtype}
\usepackage{graphicx}
\usepackage{booktabs}
\usepackage{adjustbox}
\usepackage{makecell}
\usepackage{multirow}
\usepackage{float}
\usepackage{caption}
\usepackage{subcaption}
\usepackage{hyperref}
\usepackage[accepted]{icml2024}
\usepackage{amsmath}
\usepackage{amssymb}
\usepackage{mathtools}
\usepackage{amsthm}
\usepackage[capitalize,noabbrev]{cleveref}
\usepackage[textsize=tiny]{todonotes}

\theoremstyle{plain}

\theoremstyle{definition}

\theoremstyle{remark}

\newcommand \ignore[1]{}
\newcommand{\ourmethod}{{VAE-QA}}

\icmltitlerunning{Assessing Image Quality Using a Simple Generative Representation}

\begin{document}

\twocolumn[
\icmltitle{Assessing Image Quality Using a Simple Generative Representation}

\icmlsetsymbol{equal}{*}

\begin{icmlauthorlist}
\icmlauthor{Simon Raviv}{bar_ilan}
\icmlauthor{Gal Chechik}{bar_ilan,nvidia}
\end{icmlauthorlist}

\icmlaffiliation{bar_ilan}{Department of Computer Science, Bar-Ilan University, Ramat-Gan, Israel}
\icmlaffiliation{nvidia}{NVIDIA, Tel-Aviv, Israel}

\icmlcorrespondingauthor{Simon Raviv}{simon1raviv@gmail.com}
\icmlcorrespondingauthor{Gal Chechik}{gal.chechik@gmail.com}

\icmlkeywords{IQA, Image Quality Assessment, Generative Models, VAE}

\vskip 0.2in
]

\printAffiliationsAndNotice{}

\begin{abstract}
    Perceptual image quality assessment (IQA) is the task of predicting the visual quality of an image as perceived by a human observer. Current state-of-the-art techniques are based on deep representations trained in discriminative manner. Such representations may ignore visually important features, if they are not predictive of class labels. Recent generative models successfully learn low-dimensional representations using auto-encoding and have been argued to preserve better visual features. Here we leverage existing auto-encoders and propose \ourmethod{}, a simple and efficient method for predicting image quality in the presence of a full-reference. We evaluate our approach on four standard benchmarks and find that it significantly improves generalization across datasets, has fewer trainable parameters, a smaller memory footprint and faster run time.
\end{abstract}

\section{Introduction}
\label{sec:introduction}

Assessing the visual quality of an image is a key problem with applications in numerous computer vision fields, from image restoration and enhancement to generative text-to-image models.
Current state-of-the-art methods for \textit{Image Quality Assessment} (IQA), are built on top of deep representations trained for discriminative tasks~\cite{zhang2018unreasonable,kim2017deepqa,prashnani2018pieapp,cheon2021perceptual,lao2022attentions}, taking distorted images and predicting human quality judgment.

Broadly speaking, these deep discriminative models predict perceived image quality well, but suffer from various drawbacks. They require labeled data for training the representation, they tend to be heavy and complex, and, most importantly, their discriminative representations may remove features that may be predictive about image quality, but not about class labels. Finally, discriminative representations also tend to generalize poorly to data from a different distribution~\cite{zhang2018unreasonable,bosse2018deep,prashnani2018pieapp,ding2020image,lao2022attentions}. It therefore remains a hard problem to train models that predict image quality in a way that generalizes to new datasets.

Unlike discriminative representations, recent approaches to image generation learn representations that preserve fine image content~\cite{rombach2022highresolution}. These representations can be trained self-supervised without class labels, and presumably preserve all information about image content, which may be removed by discriminative representations~\cite{li2023dreamteacher}.

Here we propose a simple approach for predicting image quality based on a \textit{Variational Auto Encoder} (VAE) generative model. Given a pre-trained VAE representation, we learn how to use its latent activation for predicting human judgment of image quality. Our approach, which we name \ourmethod{}, has significantly fewer trainable parameters and a smaller memory footprint, easily fitting on a standard consumer GPU. It also achieves state-of-the-art prediction accuracy on standard benchmarks in the field, both on new images from the same distribution and when generalizing to new datasets.

In summary, this paper makes the following contributions: \textbf{(1) }We put forward the idea that predicting image quality would be superior using a representation learned in a generative way.
\textbf{(2)} A new architecture for predicting image quality built on top of a VAE model, which learns to align features from several different layers of the VAE.
\textbf{(3)} Standardization of evaluation protocol. First, we release a data split so that future papers can compare data consistently. Second, we analyze aspects of the inference protocol, like the effect of the number of crops on prediction accuracy. 
\textbf{(4)} New SoTA results for cross-dataset generalization.

\section{Related Work}
\label{sec:related_work}

\subsection{Learning-based IQA}
\label{sec:related_work_learning_based_iqa}

Various approaches were proposed for using deep representations for IQA. The most known approach is \textit{Learned Perceptual Image Patch Similarity} (LPIPS)~\cite{zhang2018unreasonable}. They were the first to highlight the potential of using representations learned during a classification task on ImageNet~\cite{russakovsky2015imagenet} for quality prediction. DeepQA~\cite{kim2017deepqa} uses CNN to predict a visual sensitivity map that weights pixel importance in distorted images, aligning with human subjective opinions. WaDIQaM~\cite{bosse2018deep} introduces a comprehensive end-to-end deep neural network that allows for simultaneous local quality and local weight learning. Different from these methods, PieAPP~\cite{prashnani2018pieapp} focuses on learning to rank. This means the network is trained to understand the probability of one image being preferred over another.

Recent advancements in deep learning-based IQA methods have led to significant improvements in the field. ~\cite{cheon2021perceptual} proposed an \textit{Image Quality Transformer} (IQT) that applies a transformer architecture to a perceptual full-reference IQA task. The proposed model extracts perceptual feature representations from each input image using a CNN backbone. It feeds the extracted feature maps into the transformer encoder and decoder to compare reference and distorted images and predict the final score. \cite{lao2022attentions} proposed an \textit{Attention-based Hybrid Image Quality Assessment} (AHIQ) network that uses hybrid image representations learned by CNN and transformer-based networks. To predict the final score, the final image representation is learned through simple attention mechanisms.

\subsection{Non-learned IQA}
\label{sec:non_learned_iqa}

Besides learning-based IQA methods, it is common to evaluate image quality using predefined methods, including \textit{Peak Signal-to-Noise Ratio} (PSNR), \textit{Structural Similarity Index Measure} (SSIM)~\cite{wang2004image}, and \textit{Feature Similarity Index Measure }(FSIM)~\cite{zhang2011fsim}. These methods are easy to use and fast to compute, but their predictive power is very limited compared with learned approaches~\cite{zhang2018unreasonable,lao2022attentions}. It remains a challenge to simplify training of deep IQA predictors while maintaining prediction quality.

\section{Background}
\label{sec:background}

To keep this paper self-contained, we describe the concepts of generative models, \textit{Variational Auto Encoder} (VAE)~\cite{kingma2013autoencodingvb} and the various setups used for image quality assessment.

\subsection{Generative Models}
\label{sec:related_work_generative_models}

Generative models learn the underlying data distribution. The most common generative models today are diffusion models~\cite{ho2020denoising,song2022denoising,rombach2021highresolution}. Diffusion models add noise through diffusion, then de-noise their data by removing noise through denoising networks. The diffusion process is reversed to generate data. A denoising network iteratively refines a simple noise distribution, effectively 'undoing' the diffusion, until it finds a sample that closely matches the learned data distribution. As a result of this process, diffusion models generate diverse and high-quality samples that capture the complexities of the training data.

In recent years, diffusion models have shown impressive results in image generation tasks and are considered state-of-the-art. The \textit{Latent Diffusion Model} (LDM)~\cite{rombach2021highresolution} is a diffusion model that uses a latent representation of the data instead of pixels. The latent representation is learned by a \textit{Variational Auto Encoder} (VAE)~\cite{kingma2013autoencodingvb} network that maps the raw samples to a latent space. The main advantage of the LDM is its run-time efficiency, as it operates on the latent space, which is much smaller than the pixel space.

\subsection{Variational Auto Encoders}
\label{sec:background_vae}

A VAE is a type of generative model~\cite{cai2017multistage,razavi2019generating,luhman2023high} that learns to represent data in a lower-dimensional latent space. It does this by encoding input data into a latent space and then decoding or reconstructing it from this latent space. The VAE is trained to minimize the difference between the input data and the reconstructed data. The objective function of VAEs is the \textit{Evidence Lower BOund} (ELBO), which includes two components: reconstruction loss and regularization terms such as KL divergence. The reconstruction loss measures the fidelity of the reconstructed data to the original input data, while the KL divergence acts as a regularization term, encouraging the learned latent distribution to be close to a prior distribution. This balance between reconstruction and regularization allows the VAE to effectively learn the underlying data distribution.

\subsection{IQA setups}
\label{sec:background_iqa_setups}

Work in the field of IQA considers three setups: full-reference (FR-IQA) -- comparing the quality of a distorted image to a reference clean image, reduced-reference (RF-IQA) -- assessing the quality of a distorted image using some information from the reference image, and no-reference (NR-IQA) -- evaluating the distorted image is evaluated without any reference image.

\section{Our Method}
\label{sec:our_method}

\begin{figure*}[h]
    \vskip 0.2in \centering
    \includegraphics[width=\textwidth]{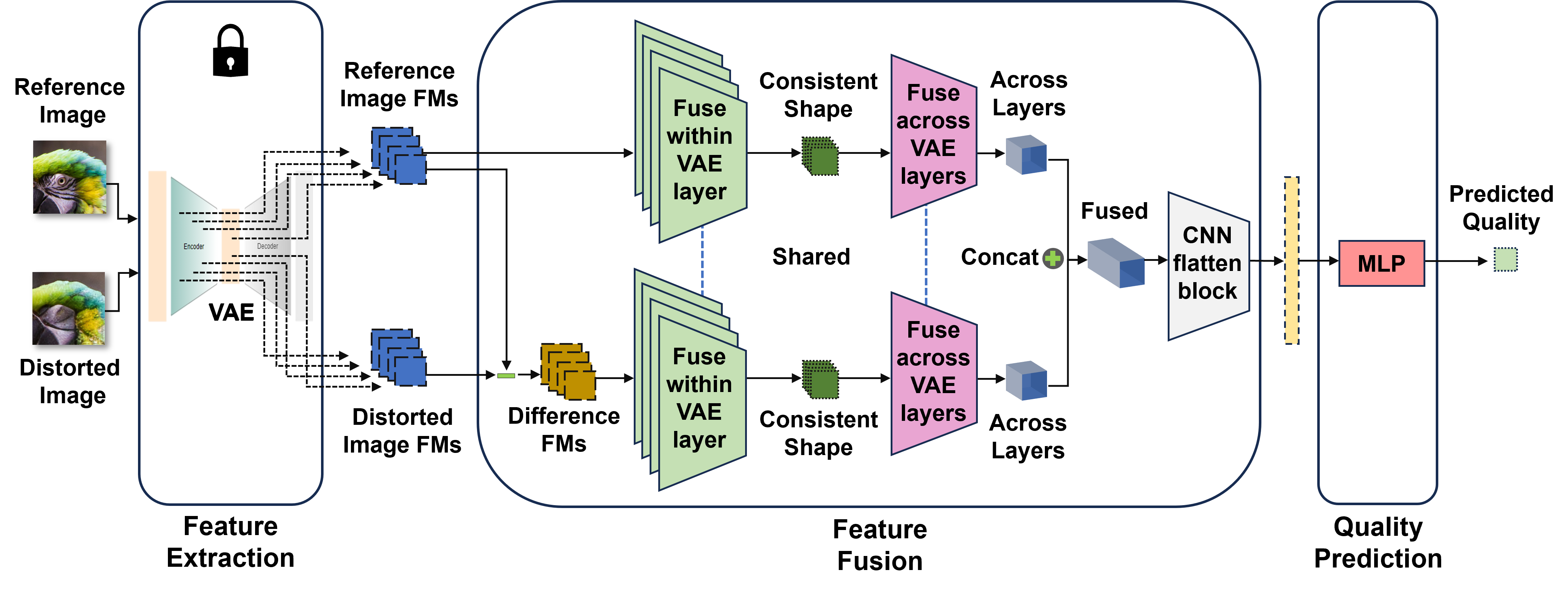}
    \caption{\ourmethod{} architecture: \textbf{Feature extraction module} extracts image representations from input images using a VAE. \textbf{Feature fusion module} combines the extracted image representations to form a compressed representation using within \& across VAE layer(s) components. \textbf{Quality prediction module} uses the compressed representation to predict the quality score of the input images using a MLP network.}
    \label{fig:vaeqa_arch}
    \vskip -0.2in
\end{figure*}

We present a model we call \textit{Variational Auto Encoder Quality Assessment} (\ourmethod{}). 
It is a deep neural network that learns to predict the quality of a distorted image, given the original reference image (Full-Reference). 

Our key idea is to use a pre-trained VAE to compute generative representations of the original image and the distorted image. The deep network is trained to predict image quality by taking as input representations from multiple layers of VAE.

FR-IQA methods typically have three main components: feature extraction, feature fusion, and quality prediction. We describe these three components of our method below. \cref{fig:vaeqa_arch} illustrates the architecture of our \ourmethod{}. A detailed description of the method's components can be found in the supplementary material.

\subsection{Feature Extraction Module}
\label{sec:our_method_feature_extraction_module}

The feature extraction module extracts image representations from input images using a VAE. Specifically, we use the VAE encoder from LDM~\cite{rombach2022highresolution} to get image representations.

By encoding a reference image and a distorted image with the VAE encoder, several feature maps are obtained from its intermediate layers. Six feature maps were chosen, spanning a broad spectrum of abstractions and layers, to encode various levels of image information.

The first two layers (image, enc1) capture basic image features in the input spatial dimension. The next three layers (enc4, enc7, enc12) are from the down-sampling part of the encoder, which extracts high-level features in reduced spatial dimensions. The next layer (enc15) is from the middle part of the encoder, which further processes the encoded representation and adds more complex features in the same spatial dimension. The last layer (enc19) is the quantization part of the encoder, which represents the latent encoded representation of the input image. These layers capture features at different abstraction levels and spatial dimensions.

\subsection{Feature Fusion Module}
\label{sec:our_method_feature_fusion_module}

The feature-fusion module is the core of the architecture. It learns to combine features extracted from the VAE into a single unified representation.
In \ourmethod{}, it has three components: Fusion within VAE layer, fusion across VAE layers, and CNN flattening block. First, a difference feature map is created by subtracting the reference feature map from the distorted feature map. Then, those feature maps are passed to the fusion within VAE layer component.

\textbf{Fusion within VAE layer.} This component learns a joint feature representation combining features within the same VAE layer, taking into account its spatial structure. We use a CNN with ReLU activation and average pooling layers, followed by group normalization over channels~\cite{wu2018group}. All feature maps regardless of layer and input dimension are mapped to an output feature dimension of $\mathbb{R}^{\text{(L+1)*128} \times 16 \times 16}$, where L is the number of layers from the VAE, and we also include the original image. This mapping allows us to combine features across layers in the next step.

\textbf{Fusion across VAE layers.} This component learns a joint feature representation combining features across different VAE layers, taking into account information from different spatial dimensions. We use a CNN with ReLU activation and average pooling layers, followed by group normalization over channels. We get a consistent shape feature map with a dimension of $\mathbb{R}^{\text{1024} \times 16 \times 16}$. Finally, we concatenate reference and difference feature maps to form a single feature map with a dimension of $\mathbb{R}^{\text{2048} \times 16 \times 16}$.

\begin{table*}[h]
    \caption{ IQA datasets for model training and performance evaluation.}
    \label{tab:datasets}
    \vskip 0.15in \centering\small\sc
    \begin{tabular}{@{}lccccc@{}}
        \toprule
        \multicolumn{1}{c}{\textbf{\makecell[c]{Dataset}}} & \multicolumn{1}{c}{\textbf{\makecell[c]{\# Reference\\images}}} & \multicolumn{1}{c}{\textbf{\makecell[c]{\# Distorted\\images}}} & \multicolumn{1}{c}{\textbf{\makecell[c]{\# Distortion\\type}}} & \multicolumn{1}{c}{\textbf{\makecell[c]{\# Rating}}} & \multicolumn{1}{c}{\textbf{\makecell[c]{Rating\\type}}} \\
        \midrule
        LIVE~\cite{sheikh2006statistical}   & 29 & 779    & 5  & 25K   & DMOS \\
        CSIQ~\cite{larson2010most}          & 30 & 866    & 6  & 5K    & DMOS \\
        TID2013~\cite{ponomarenko2015image} & 25 & 3,000  & 24 & 500K  & MOS  \\
        KADID-10k~\cite{lin2019kadid}       & 81 & 10,125 & 25 & 30.4K & DMOS \\
        \bottomrule
    \end{tabular}
    \vskip -0.1in
\end{table*}

\textbf{CNN flatten block.} This component transforms a 2D features map into a 1D vector, used for quality prediction. We use a CNN with ReLU activation and average pooling layers to compress the feature maps. Features are also normalized using group normalization. Finally, we flatten the feature maps to form a compressed representation vector with a dimension of $\mathbb{R}^{\text{4096}}$.

\subsection{Quality Prediction Module}
\label{sec:our_method_quality_prediction_module}

The quality prediction module is a 3 layer MLP network. It takes the fused representation vector as input and predicts the quality score of the input images.

\section{Experiments}
\label{sec:experiments}

\subsection{Compared Methods}
\label{sec:experiments_compared_methods}

We compared \ourmethod{} with four recent baselines that use deep representations and achieve high prediction accuracy.
(1) LPIPS~\cite{zhang2018unreasonable} A pioneering paper that showed the advantage of using deep representations to predict perceptual image quality.
(2) DeepQA~\cite{kim2017deepqa} predicts a visual sensitivity map that weights pixel importance in distorted images, aligning with human subjective opinions.
(3) PieAPP~\cite{prashnani2018pieapp}, trained using ranking loss.
(4) AHIQ~\cite{lao2022attentions}, added features from vision transformers to standard CNN features.

Notably, ~\cite{seo2020novel} showed that quality predictions can be improved by predicting a saliency map designed to mimic human perception, and adding that as a side-channel to deep representations. Since this paper focuses on finding good deep features, and since we could not obtain their code for comparisons, we do not report their results in the main table, but discuss the effect of salience maps in the supplementary.

We also report standard non-deep predictors, like PSNR and SSIM~\cite{wang2004image}, since these are commonly used.

\begin{table*}[ht]
    \caption{\textbf{Within-dataset evaluation:} Performance scores on three standard IQA datasets: LIVE, CSIQ, and TID2013. Scores for our method are averages across three seeds, the standard error is the order of 0.01, and is reported in the supplementary for clarity.
        Scores for~\cite{lao2022attentions} and~\cite{zhang2018unreasonable} were reproduced using authors code and are marked by (*), see also \cref{tab:protocol_effect}.
        Scores for all other methods were taken from original papers.}
    \label{tab:td_eval}
    \vskip 0.1in \centering\small\sc
    \begin{tabular}{@{}lcc|cc|cc|cc@{}}
        \toprule
        \multicolumn{1}{c}{\textbf{\makecell[l]{Method}}} & \multicolumn{2}{c}{\textbf{ LIVE}} & \multicolumn{2}{c}{\textbf{ CSIQ}} & \multicolumn{2}{c}{\textbf{TID2013}} & \multicolumn{2}{c}{\textbf{ Overall}} \\
        \cmidrule(lr){2-3} \cmidrule(lr){4-5} \cmidrule(lr){6-7} \cmidrule(lr){8-9}
        & \makecell[c]{PLCC} & \makecell[c]{SRCC} & \makecell[c]{PLCC} & \makecell[c]{SRCC} & \makecell[c]{PLCC} & \makecell[c]{SRCC} & \makecell[c]{PLCC} & \makecell[c]{SRCC} \\
        \midrule
        PSNR                             & .865 & .873 & .819 & .810 & .677 & .687 & .787 & .790 \\
        SSIM~\cite{wang2004image}        & .937 & .948 & .852 & .865 & .777 & .727 & .855 & .847 \\
        FSIMc~\cite{zhang2011fsim}       & .961 & .965 & .919 & .931 & .877 & .851 & .919 & .916 \\
        VSI~\cite{zhang2014vsi}          & .948 & .952 & .928 & .942 & .900 & .897 & .925 & .930 \\
        NLPD~\cite{valero2016perceptual} & .932 & .937 & .923 & .932 & .839 & .800 & .898 & .890 \\
        GMSD~\cite{xue2014gradient}      & .957 & .960 & .945 & .950 & .855 & .804 & .919 & .905 \\
        SCQI~\cite{bae2016scqi}          & .937 & .948 & .927 & .943 & .907 & .905 & .924 & .932 \\
        \midrule
        \textbf{Deep methods} & & & & & & \\
        LPIPS~\cite{zhang2018unreasonable} (*) & .823 & .911 & .873          & .934          & .758 & .779 & .818 & .875 \\
        DeepQA~\cite{kim2017deepqa}            & .982 & .981 & .965          & .961          & .947 & .939 & .818 & .875 \\
        PieAPP~\cite{prashnani2018pieapp}      & .986 & .977 & \textbf{.975} & \textbf{.973} & .946 & .945 & .969 & .965 \\
        AHIQ~\cite{lao2022attentions} (*) & \textbf{.988} & \textbf{.985} & .971          & .968 & .927          & .922          & .962          & .958          \\
        \ourmethod{} (ours)               & .984          & .981          & \textbf{.974} & .968 & \textbf{.961} & \textbf{.958} & \textbf{.973} & \textbf{.969} \\
        \bottomrule
    \end{tabular}
    \vskip -0.1in
\end{table*}
\begin{figure*}[ht]
    \vskip 0.1in \centering
    \begin{subfigure}[b]{0.32\textwidth}
        \centering
        \includegraphics[width=\textwidth]{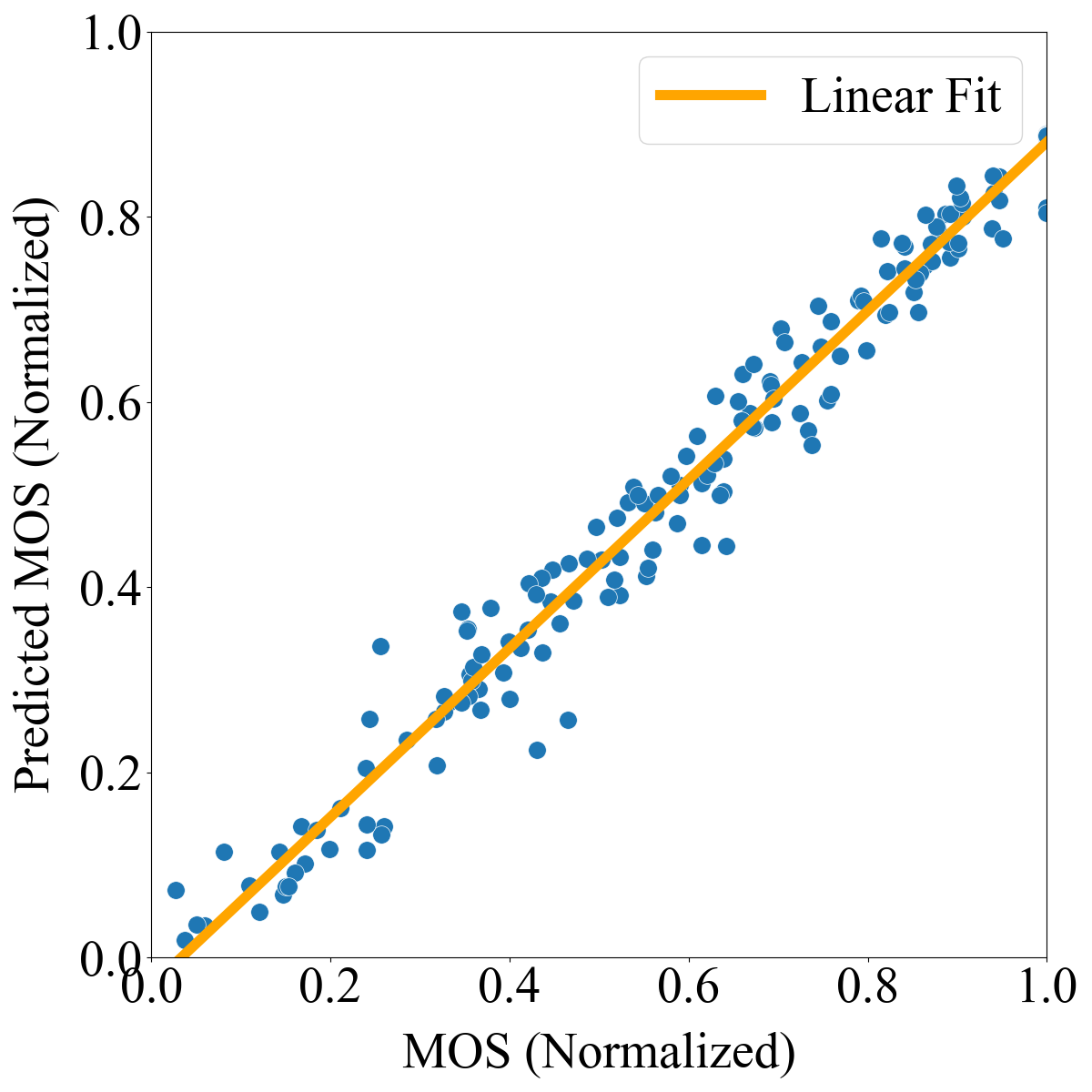}
        \caption{LIVE dataset}
        \label{fig:live_val_split_labels_scatter_plot}
    \end{subfigure}
    \hfill
    \begin{subfigure}[b]{0.32\textwidth}
        \centering
        \includegraphics[width=\textwidth]{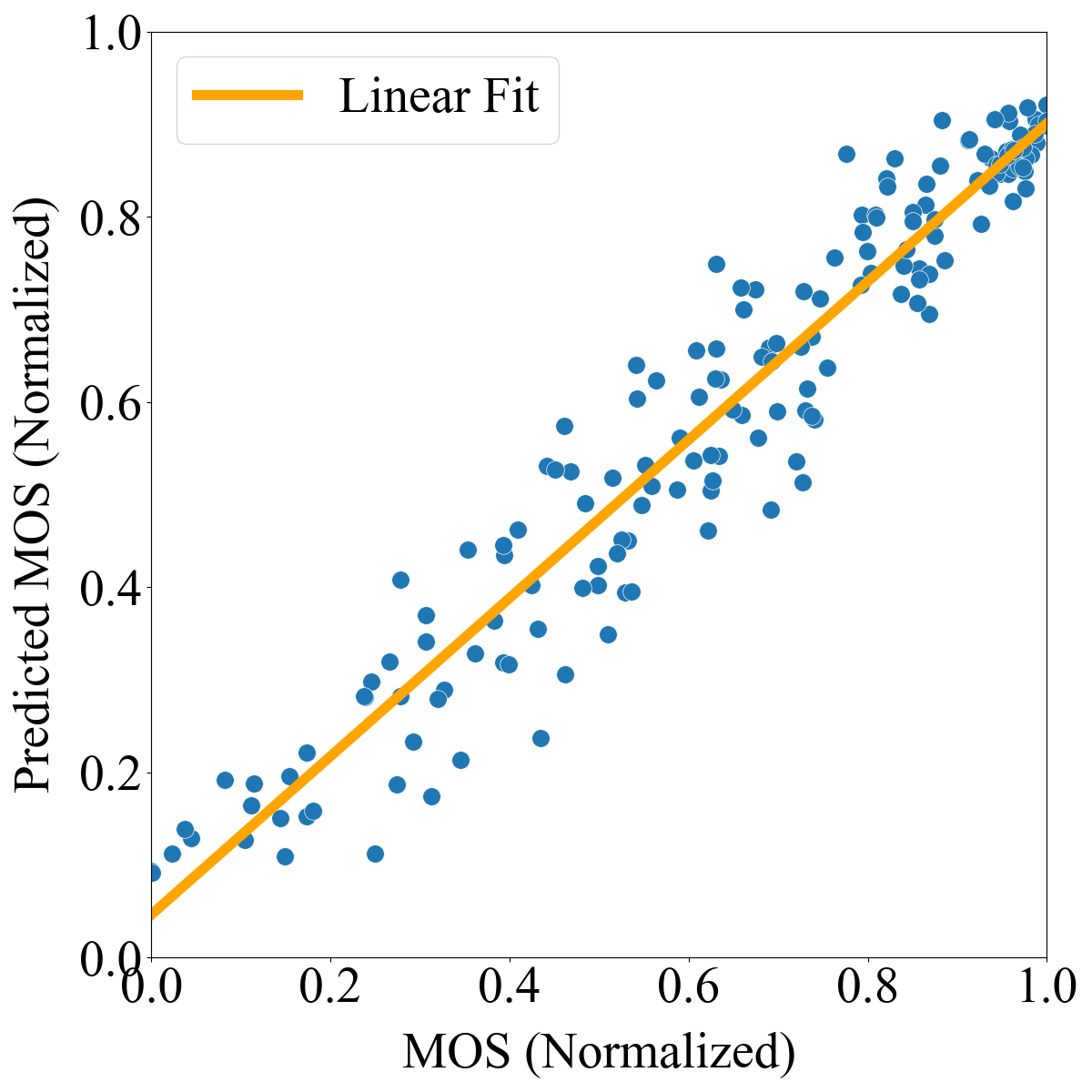}
        \caption{CSIQ dataset}
        \label{fig:csiq_val_split_labels_scatter_plot}
    \end{subfigure}
    \hfill
    \begin{subfigure}[b]{0.32\textwidth}
        \centering
        \includegraphics[width=\textwidth]{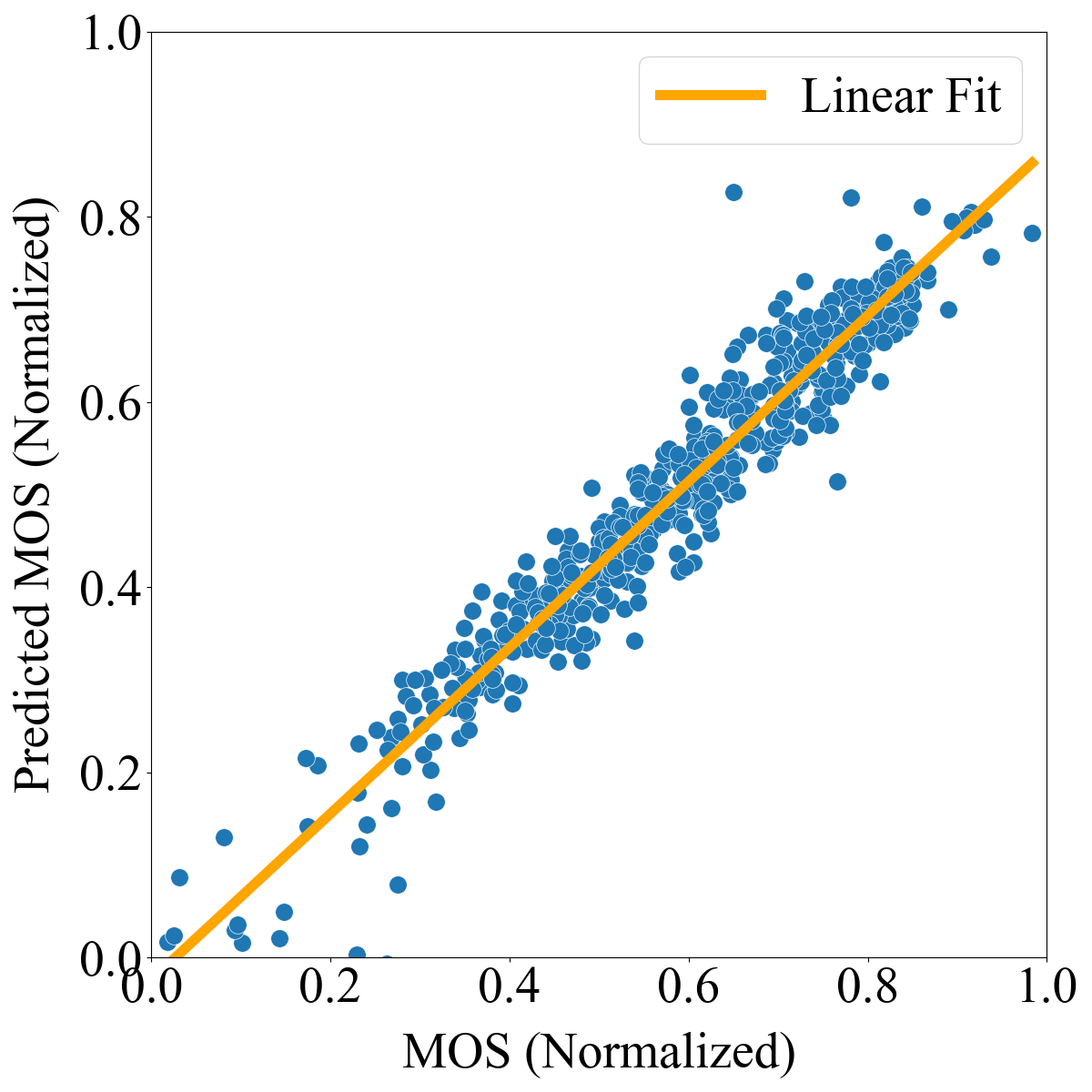}
        \caption{TID2013 dataset}
        \label{fig:tid2013_val_split_labels_scatter_plot}
    \end{subfigure}
    \caption{MOS vs. Predicted MOS for three IQA datasets.}
    \label{fig:td_eval_mos_vs_predicted_mos}
\vskip -0.1in
\end{figure*}

\begin{table*}[ht]
    \caption{\textbf{Cross dataset evaluation}. Trained on KADID-10k dataset and tested on LIVE, CSIQ, and TID2013 datasets. Performance scores for all methods except of AHIQ were taken from their papers. (*) stands for results that we reproduced using the authors code. (**) stands for results borrowed from~\cite{ding2020image}.}
    \label{tab:cd_eval_kadid10K}
    \vskip 0.15in \centering\small\sc
    \begin{tabular}{@{}lcc|cc|cc|cc@{}}
        \toprule
        \multicolumn{1}{c}{\textbf{\makecell[l]{Method}}} & \multicolumn{2}{c}{\textbf{{KADID} $\rightarrow$ LIVE}} & \multicolumn{2}{c}{\textbf{KADID$ \rightarrow$ CSIQ}} & \multicolumn{2}{c}{\textbf{KADID$ \rightarrow$ TID2013}} & \multicolumn{2}{c}{\textbf{Overall}}                                                                                     \\
        \cmidrule(lr){2-3} \cmidrule(lr){4-5} \cmidrule(lr){6-7} \cmidrule(lr){8-9}
        & \makecell[c]{PLCC} & \makecell[c]{SRCC} & \makecell[c]{PLCC} & \makecell[c]{SRCC} & \makecell[c]{PLCC} & \makecell[c]{SRCC} & \makecell[c]{PLCC} & \makecell[c]{SRCC} \\
        \midrule
        LPIPS~\cite{zhang2018unreasonable} (**) & .934 & .932 & .896 & .876 & .749 & .670 & .860 & .826 \\
        WaDIQaM~\cite{bosse2018deep}            & .940 & .947 & .901 & .909 & .834 & .831 & .892 & .896 \\
        PiPieAPP~\cite{prashnani2018pieapp}     & .908 & .919 & .877 & .892 & .859 & .876 & .881 & .896 \\
        DISTS~\cite{ding2020image}              & .934 & .932 & .896 & .876 & .749 & .670 & .860 & .826 \\
        AHIQ~\cite{lao2022attentions} (*)       & \textbf{.956} & .954          & .945          & .940          & .891          & .888          & .931          & .927          \\
        \ourmethod{} (ours)                     & .945          & \textbf{.967} & \textbf{.952} & \textbf{.960} & \textbf{.905} & \textbf{.900} & \textbf{.934} & \textbf{.942} \\
        \bottomrule
    \end{tabular}
    \vskip 0.15in
\end{table*}
\begin{table*}[ht]
    \caption{\textbf{Cross dataset evaluation}. Trained on TID2013 dataset and tested on LIVE and CSIQ datasets. "CENTER" refers to using a single center crop. Performance scores for all methods were taken from their papers.}
    \label{tab:cd_eval_tid2013}
    \vskip 0.15in \centering\small\sc
    \begin{tabular}{@{}lcc|cc|cc@{}}
        \toprule
        \multicolumn{1}{c}{\textbf{\makecell[l]{Method}}} & \multicolumn{2}{c}{\textbf{{TID2013} $\rightarrow$ LIVE}} & \multicolumn{2}{c}{\textbf{TID2013$ \rightarrow$ CSIQ}} & \multicolumn{2}{c}{\textbf{Overall}} \\
        \cmidrule(lr){2-3} \cmidrule(lr){4-5} \cmidrule(lr){6-7}
        & \makecell[c]{PLCC} & \makecell[c]{SRCC} & \makecell[c]{PLCC} & \makecell[c]{SRCC} & \makecell[c]{PLCC} & \makecell[c]{SRCC} \\
        \midrule
        WaDIQaM~\cite{bosse2018deep}    & -             & .936          & -             & .931          & -             & .934          \\
        DOG-SSIM~\cite{pei2015dogssimc} & -             & \textbf{.948} & -             & .925          & -             & .937          \\
        \ourmethod{} (center)           & .923          & .942          & .938          & .935          & .931          & .939          \\
        \ourmethod{} (ours)             & \textbf{.930} & \textbf{.948} & \textbf{.943} & \textbf{.941} & \textbf{.937} & \textbf{.945} \\
        \bottomrule
    \end{tabular}
    \vskip -0.15in
\end{table*}
\begin{figure*}[ht]
    \centering
    \begin{subfigure}[b]{0.3\textwidth}
        \centering
        \includegraphics[width=\textwidth]{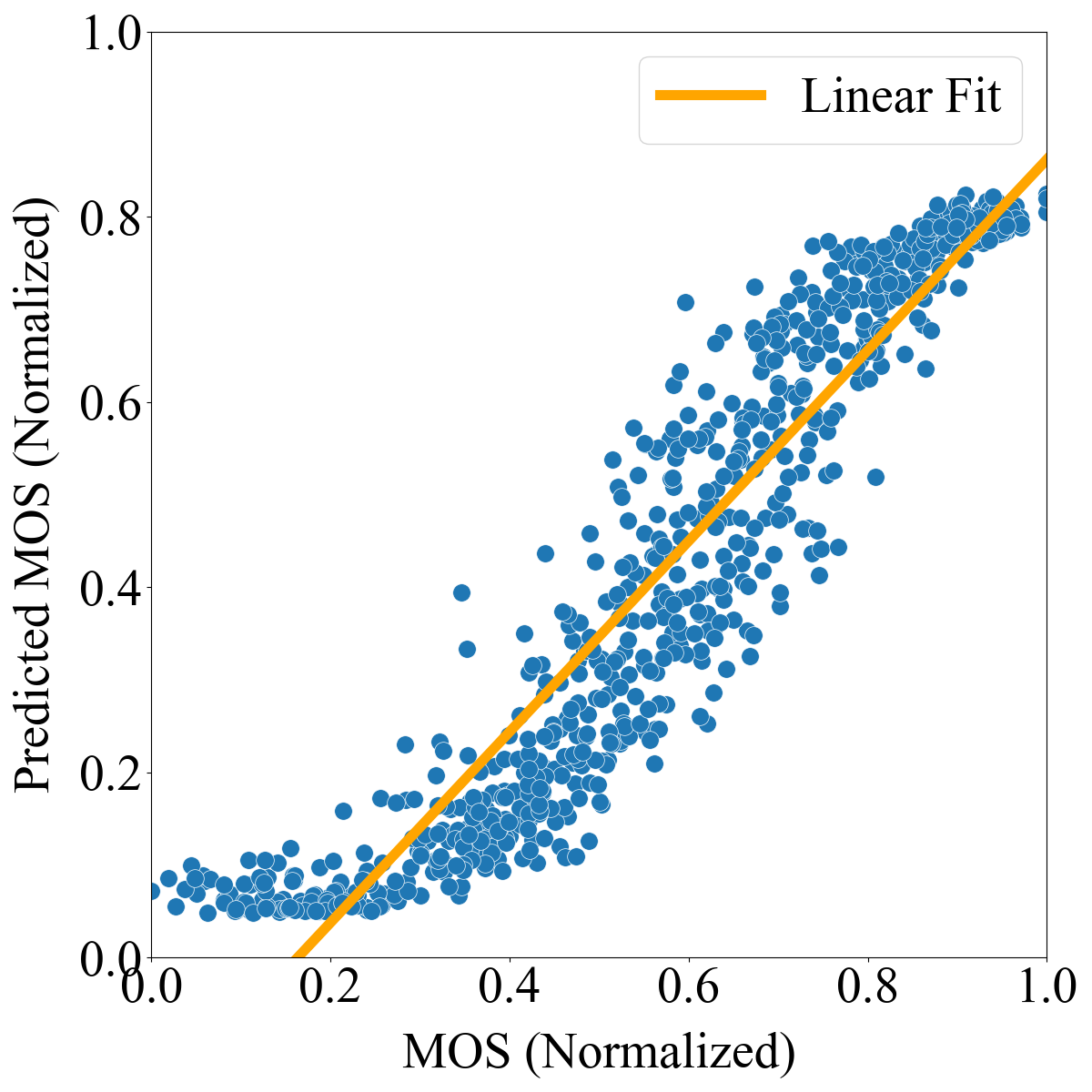}
        \caption{LIVE dataset.}
        \label{fig:live_train_kadid-10K_test_split_labels_scatter_plot}
    \end{subfigure}
    \hfill
    \begin{subfigure}[b]{0.3\textwidth}
        \centering
        \includegraphics[width=\textwidth]{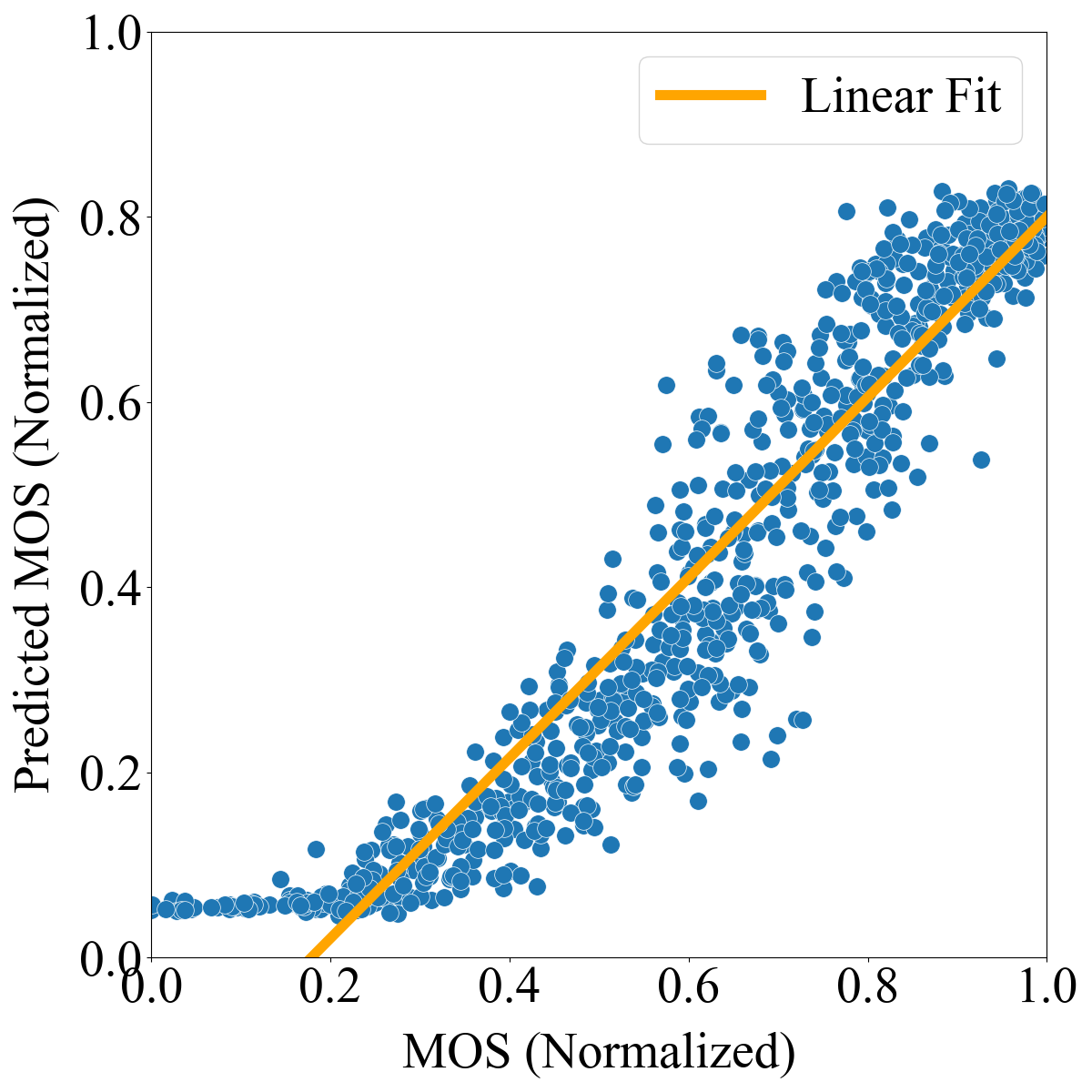}
        \caption{CSIQ dataset.}
        \label{fig:csiq_train_kadid-10K_test_split_labels_scatter_plot}
    \end{subfigure}
    \hfill
    \begin{subfigure}[b]{0.3\textwidth}
        \centering
        \includegraphics[width=\textwidth]{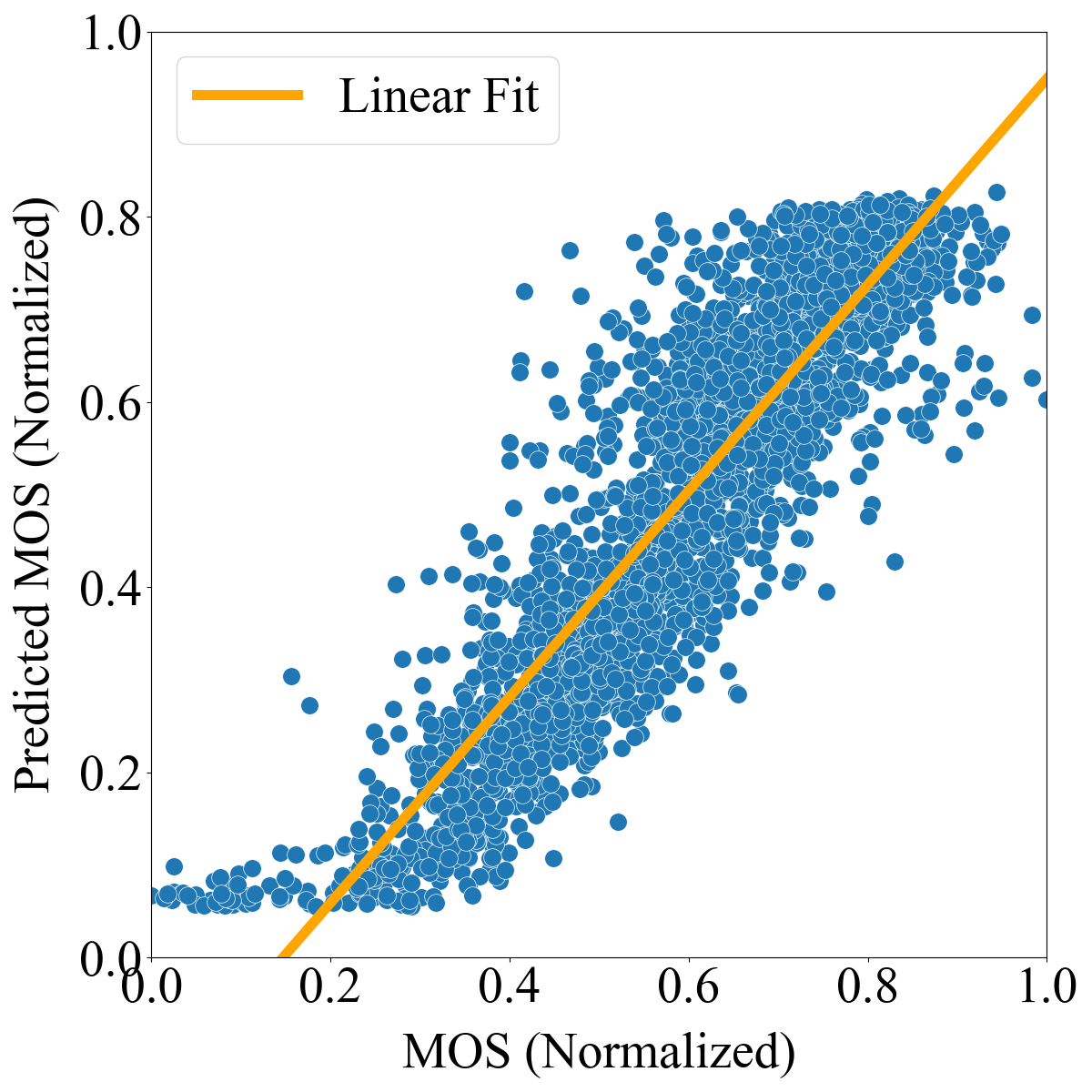}
        \caption{TID2013 dataset.}
        \label{fig:tid2013_train_kadid-10K_test_split_labels_scatter_plot}
    \end{subfigure}
    \caption{MOS vs. Predicted MOS. Trained on KADID-10k, tested on other IQA datasets.}
    \label{fig:cd_eval_mos_vs_predicted_mos}
\end{figure*}

\begin{table*}[ht]
    \caption{The effect of inference protocol on prediction quality for within-dataset experiments.}
    \label{tab:protocol_effect}
    \vskip 0.15in \centering\small\sc\scalebox{0.95}{
    \begin{tabular}{@{}lcc|cc|cc|cc@{}}
        \toprule
        \multicolumn{1}{c}{\textbf{\makecell[l]{Method}}} & \multicolumn{2}{c}{\textbf{LIVE}} & \multicolumn{2}{c}{\textbf{CSIQ}} & \multicolumn{2}{c}{\textbf{TID2013}} & \multicolumn{2}{c}{\textbf{Overall}} \\
        \cmidrule(lr){2-3} \cmidrule(lr){4-5} \cmidrule(lr){6-7} \cmidrule(lr){8-9}
        & \makecell[c]{PLCC} & \makecell[c]{SRCC} & \makecell[c]{PLCC} & \makecell[c]{SRCC} & \makecell[c]{PLCC} & \makecell[c]{SRCC} & \makecell[c]{PLCC} & \makecell[c]{SRCC} \\
        \midrule
        \textbf{20 Random crops} & & & & & & & & \\
        AHIQ~\cite{lao2022attentions} (reported) & .989 & .984 & .978 & .975 & .968 & .962 & .978 & .974 \\
        AHIQ~\cite{lao2022attentions} (reproduced) & \textbf{.988} & \textbf{.985} & .971 & \textbf{.968} & .927 & .922 & .962 & .958 \\
        \ourmethod{} (ours) & .984 & .981 & \textbf{.974} & \textbf{.968} & \textbf{.961} & \textbf{.958} & \textbf{.973} & \textbf{.969} \\
        \midrule
        \textbf{Center crop} & & & & & & & & \\
        AHIQ~\cite{lao2022attentions} (reproduced) & \textbf{.987} & \textbf{.984} & .970 & .965 & .931 & .922 & .963 & .957 \\
        \ourmethod{} (ours) & .984 & .981 & \textbf{.971} & \textbf{.966} & \textbf{.958} & \textbf{.954} & \textbf{.971} & \textbf{.967} \\
        \bottomrule
    \end{tabular}}
    \vskip -0.15in
\end{table*}
\subsection{Evaluation Protocol}
\label{sec:eval}

\subsubsection*{Datasets:}
We evaluate our approach with the four datasets commonly used in evaluations of FR-IQA methods LIVE~\cite{sheikh2006statistical}, CSIQ~\cite{larson2010most}, TID2013~\cite{ponomarenko2015image}, and KADID-10k~\cite{lin2019kadid}. \cref{tab:datasets} provides details about each dataset.

Usually, LIVE, CSIQ, and TID2013 are used for "in-distribution" evaluations. LIVE and CSIQ contain several distortion types, such as blurring, noise, and JPEG compression. TID2013, is much more comprehensive and contains 24 distortion types. These include new noise types, such as impulse noise and high frequency noise, which negatively affect image quality. Including spatial and color shifts, such as local block-wise distortions, mean shifts, and contrast changes, which alter an image's appearance. This makes TID2013 more challenging.

Recent papers~\cite{bosse2018deep,prashnani2018pieapp,ding2020image,lao2022attentions} used KADID-10k to evaluate cross-dataset generalization. KADID-10k is more than three times larger than TID2013, and includes more reference images.

\subsubsection*{Target values:}
In FR-IQA, \textit{Mean Opinion Score} (MOS) refers to the average score a group of people give an image when comparing its original version to another image. The objective of our method is to predict MOS. Details on MOS vs. \textit{Difference MOS} (DMOS) are available in the supplementary.

\subsubsection*{Data splits:}
There is no standard data split in this domain. Therefore we split each dataset randomly into training (80\%), and test (20\%) sets and make that split available at \textit{\href{https://github.com/SimonRaviv/VAE-QA/tree/main/datasets_splits}{IQA Standard Datasets Splits}}.

We repeated the split three times and reported below the average over the three random splits. Splitting is done according to the reference image, so all distortions of the same image are in the same split. Images from test data are not visible during training.

To tune the hyperparameters of all models, we further split the training set into train and validation (60\% of the original data into train and 20\% into validation).

\subsubsection*{Evaluation metrics:}
We follow the standard protocol in this field and report the correlation between predicted values and ground truth human judgment of image quality for each distorted image. Specifically, we compute \textit{Pearson Linear Correlation Coefficient} (PLCC) and the (non-linear) \textit{Spearman Rank Correlation Coefficient} (SRCC).

\subsubsection*{Inference protocol:} 
There is no single standardized evaluation protocol in this domain. Different previous papers take different approaches to how they handled images of different sizes. They use various sizes and numbers of crops from images and integrating information from crops in various ways.

For instance, JND-SalCAR~\cite{seo2020novel} used $32\times32$ patches sampled from reference and distorted images, to train and test their network. However, this approach may not fully capture the global quality of the image, as it ignores the spatial relationships between patches. Moreover, it may introduce bias due to the patch selection strategy, as some regions may be over- or under-represented. In contrast, AHIQ~\cite{lao2022attentions} used the entire image as input, and reported the average of 20 random crops from the image as the final score.

To disentangle the effect of the number of crops, we report the quality of our method using the same protocol as~\cite{lao2022attentions}, and later study the effect of various crops.

Out of all previous strong work in this area, only \cite{lao2022attentions} provided code that allowed us to reproduce the results using the same protocol.

\subsection{Implementation Details}
\label{sec:exp_implementation_details}

\subsubsection*{Pre-trained VAE:}
We use a VAE~\cite{kingma2013autoencodingvb} from LDM \cite{rombach2022highresolution} that was pre-trained on OpenImages~\cite{Kuznetsova_2020}. Specifically, the VAE variant we used was trained with KL divergence regularized latent space and a down sample factor of 8 (aka LDM-8).

\subsubsection*{Data processing:}
Input images are normalized to the range $[-1, 1]$ and cropped to $256 \times $256. During training, we randomly crop input images with a uniform distribution and flip them horizontally with a probability of 0.5.

\subsubsection*{Image representation:}
Each image was represented using 7 maps: the $256\times256$ cropped image itself and six layers of VAE encoder outputs, layers \{1, 4, 7, 12, 15, 19\}. These layers include the initial CNN layer (1), 3 down-sampling stages (4, 7, 12), a mid-stage (15), and an end part (19) that reduces channels and quantizes to the final latent space representation.

\subsubsection*{Normalization and regularization:}
The signal in the model is normalized by group normalization~\cite{wu2018group}. We use 32 groups for normalization layers. For 2D dropout~\cite{tompson2015efficient} layers in the feature fusion module, we use a dropout rate of 0.2, and for the quality prediction module dropout~\cite{hinton2012improving} layers, we apply a dropout rate of 0.3.

\subsubsection*{Optimization:}
As a training loss, we minimized the mean squared error between the predicted and true MOS labels, normalized to the range $[0, 1]$. We use Adam~\cite{kingma2017adam} optimizer with an initial learning rate of $10^{-4}$, weight decay of $10^{-5}$, and batch size of 8. Learning rate was scheduled with Cosine Annealing~\cite{loshchilov2017sgdr} with a minimum learning rate of 0 and maximum scheduler steps of 50. The model is trained for 50 epochs and the best model is selected based on the highest PLCC and SRCC evaluation metrics. Specifically, we followed the protocol by~\cite{lao2022attentions}: at each epoch, if the PLCC or SRCC was greater than the highest recorded till that epoch, the model of that epoch was considered the best model.

\subsubsection*{SW \& HW:}
The model was trained using a single NVIDIA A6000 GPU with 48GB of memory. The model is implemented over PyTorch~\cite{paszke2019pytorch} and PyTorch Lightning~\cite{falconPyTorchLightning2019} libraries.

\section{Results}
\label{sec:results}

We evaluate \ourmethod{} in two main scenarios. First, generalizing to new images from the same dataset. Then, a more realistic scenario of generalizing to images from a different dataset not seen during training. We follow standard evaluation protocols as possible.

\subsection{Within-dataset Evaluations} 
\label{sec:within_dataset}

We first evaluate the model by training and testing it on images from the same data distribution. \cref{tab:td_eval} shows the linear and non-linear correlation over the test split for each of the 3 standard datasets. For LIVE and CSIQ datasets, \ourmethod{} shows a small improvement, while it improves considerably for TID2013 and across datasets overall.

Figure \ref{fig:td_eval_mos_vs_predicted_mos} shows the relation between predicted and ground-truth quality. The scatter plot suggests this relation is largely linear in this regime.

\subsection{Cross dataset Evaluations} 
\label{sec:cross_dataset}

To evaluate the generalization performance of the model, we trained the model on the entire KADID-10k dataset and evaluated it on the full set of LIVE, CSIQ, and TID2013 datasets. \cref{tab:cd_eval_kadid10K} report the average PLCC and SRCC values of this experiment. \ourmethod{} consistently improves over current methods.

We further tested generalization by training the model on the TID2013 dataset and testing on the LIVE and CSIQ datasets. \cref{tab:cd_eval_tid2013} shows the results of this experiment. Our model generalizes robustly to new data distributions.

Figure \ref{fig:cd_eval_mos_vs_predicted_mos} shows the relation between predicted MOS and ground truth for the cross-dataset experiments.

\subsection{Analysis}
\label{sec:ablation_study}

We explore various parameters that affect the quality of the results including the evaluation protocol and number of crops, and the variance across crops and seeds.

\subsubsection*{The effect of evaluation protocol:}
Different studies in this field~\cite{seo2020novel,prashnani2018pieapp,lao2022attentions} use different ways to preprocess a given image into a square input to the model. \cref{tab:protocol_effect} and \cref{tab:cd_eval_kadid10K_protocol} reports the correlations observed when using 20 crops compared with a single central crop. It also shows results obtained from the original AHIQ paper and the correlations obtained where we ran the authors' code in our system. The authors have been extremely helpful, and we made an effort to reproduce all hyperparameters in the model. The differences can be attributed to "lucky" random seeds, or to hyperparameter differences that we could not trace.

\begin{figure*}[ht]
    \vskip 0.2in \centering
    \includegraphics[width=\textwidth]{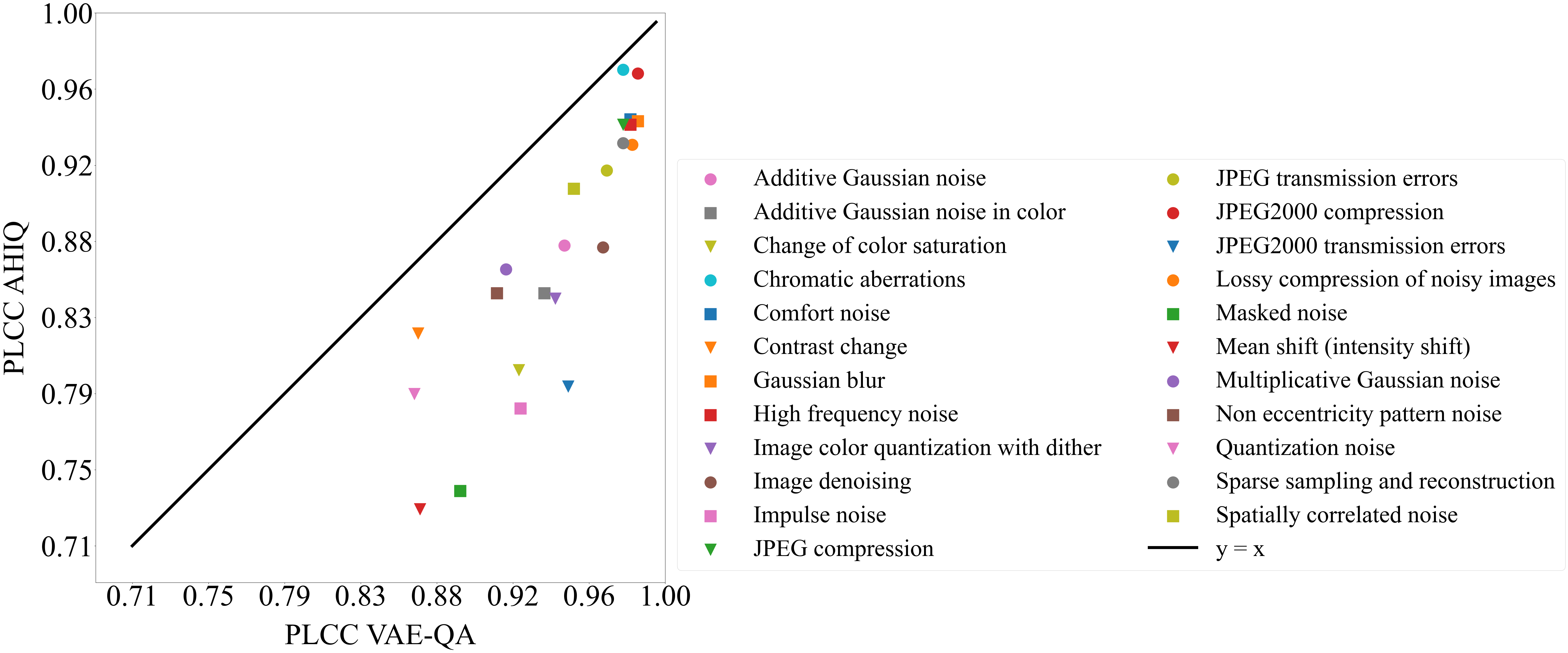}
    \caption{Quality prediction by distortion type on TID2013 dataset. The figure compares PLCC obtained with our \ourmethod{} and AHIQ.}
    \label{fig:tid2013_distortion_types}
    \vskip -0.2in
\end{figure*}

\begin{table*}[ht]
    \caption{The effect of inference protocol on cross-dataset generalization.}
    \label{tab:cd_eval_kadid10K_protocol}
    \vskip 0.15in \centering\small\sc
    \begin{tabular}{@{}lcc|cc|cc|cc@{}}
        \toprule
        \multicolumn{1}{c}{\textbf{\makecell[l]{Method}}} & \multicolumn{2}{c}{\textbf{KADID$\rightarrow$LIVE}} & \multicolumn{2}{c}{\textbf{KADID$\rightarrow$CSIQ}} & \multicolumn{2}{c}{\textbf{KADID$\rightarrow$TID2013}} & \multicolumn{2}{c}{\textbf{Overall}} \\
        \cmidrule(lr){2-3} \cmidrule(lr){4-5} \cmidrule(lr){6-7} \cmidrule(lr){8-9}
        & \makecell[c]{PLCC} & \makecell[c]{SRCC} & \makecell[c]{PLCC} & \makecell[c]{SRCC} & \makecell[c]{PLCC} & \makecell[c]{SRCC} & \makecell[c]{PLCC} & \makecell[c]{SRCC} \\
        \midrule
        \textbf{Center crop} \\
        AHIQ~\cite{lao2022attentions} (reproduced) & \textbf{.949} & .942 & .930 & .922 & .868 & .862 & .916 & .909 \\
        \ourmethod{} (ours) & .942 & \textbf{.964} & \textbf{.949} & \textbf{.957} & \textbf{.897} & \textbf{.893} & \textbf{.929} & \textbf{.938} \\
        \midrule
        \textbf{20 random crops } \\
        AHIQ~\cite{lao2022attentions} (reported) & .952 & .970 & .955 & .951 & .899 & .901 & .935 & .941 \\
        AHIQ~\cite{lao2022attentions} (reproduced) & \textbf{.956} & .954 & .945 & .940 & .891 & .888 & .931 & .927 \\
        \ourmethod{} (ours) & .945 & \textbf{.967} & \textbf{.952} & \textbf{.960} & \textbf{.905} & \textbf{.900} & \textbf{.934} & \textbf{.942} \\
        \bottomrule
    \end{tabular}
    \vskip -0.15in
\end{table*}

\begin{table*}[ht]
    \caption{The effect of the variance across crops and seeds on prediction quality.}
    \label{tab:performance_across_crops_seeds}
    \vskip 0.15in \centering\small\sc
    \begin{tabular}{@{}lcc|cc|cc@{}}
        \toprule
        \multicolumn{1}{c}{\textbf{\makecell[c]{Method}}} & \multicolumn{2}{c}{\textbf{LIVE}} & \multicolumn{2}{c}{\textbf{CSIQ}} & \multicolumn{2}{c}{\textbf{TID2013}} \\
        \cmidrule(lr){2-3} \cmidrule(lr){4-5} \cmidrule(lr){6-7}
        & \makecell[c]{PLCC} & \makecell[c]{SRCC} & \makecell[c]{PLCC} & \makecell[c]{SRCC} & \makecell[c]{PLCC} & \makecell[c]{SRCC} \\
        \midrule
        AHIQ~\cite{lao2022attentions} (reproduced) & .0023$\pm$.0008     & .0023$\pm$.0007      & .0030$\pm$.0012 & .0026$\pm$.0005 & .0052$\pm$.0012  & .0068$\pm$.0015 \\
        \ourmethod{} (ours)                        & .0023$\pm$.0009 & .0020$\pm$.0009  & .0025$\pm$.0014 & .0027$\pm$.0007 & .0018$\pm$.0008  & .0023$\pm$.0008 \\
        \bottomrule
    \end{tabular}
    \vskip -0.15in
\end{table*}

\subsubsection*{The effect of the number of crops:}
\begin{figure}[H]
    \centering    
    \includegraphics[width=0.85\linewidth]{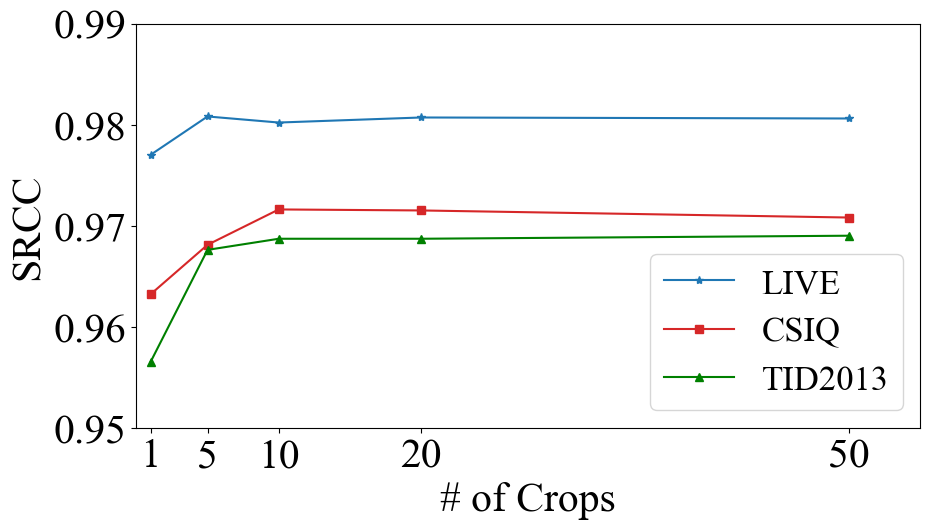}
    \caption{The effect of the number of crops on the SRCC.}
    \label{fig:num_crops_vs_srcc}
\end{figure}

Correlation results were also examined in relation to the number of crops. See \cref{fig:num_crops_vs_srcc}. At 20 crops, saturation is noticeable.

\subsubsection*{Variance across crops and seeds:}
\cref{tab:performance_across_crops_seeds} examines the variance across crops and seeds together. It shows the standard deviation over 20 crops, averaged across 3 seeds, and the standard deviation across those three seeds.

We observe a small variance between seeds, and a larger variance between crops. This is consistent with the results in \cref{fig:num_crops_vs_srcc}.

\subsubsection*{Correlation by distortion type}
To obtain more insight into the performance of \ourmethod{}, we measured the quality of quality prediction for different types of distortion. \cref{fig:tid2013_distortion_types} compares the prediction accuracy measured using PLCC, for our method and AHIQ, the current SoTA. 
The VAE representation underlying \ourmethod{}, appears to help in all distortion types, and provides on average larger improvements for those types that are challenging for AHIQ (masked noise and intensity shift).

\begin{table*}[ht]
    \caption{Number of parameters and model size.}
    \label{tab:mem_consumption}
    \vskip 0.15in \centering\small\sc
    \begin{tabular}{@{}lcc|c|c@{}}
        \toprule
        \multirow{2}{*}{\textbf{\makecell[c]{Method}}} & \multicolumn{1}{c}{\textbf{\makecell[c]{Non\\trainable [M]}}} & \multicolumn{1}{c}{\textbf{Trainable [M]}} & \multicolumn{1}{c}{\textbf{\makecell{Total\\params [M]}}} & \multicolumn{1}{c}{\textbf{\makecell{Params\\size [MB]}}} \\
        \midrule
        AHIQ~\cite{lao2022attentions} & 112.1 & 27.2  & 139.3 & 557.2 \\
        \ourmethod{} (ours)           & 34.2  & 14.4  & 48.5  & 194.2 \\
        \midrule
        \textbf{Diff [\%]}            & 69.49 & 47.06 & 65.18 & 65.15 \\
        \bottomrule
    \end{tabular}
    \vskip -0.15in
\end{table*}

\subsection{Runtime and Memory Footprint}
\label{sec:results_runtime_memory_footprint}

\begin{figure}[H]
    \vskip 0.2in \centering
    \begin{subfigure}[b]{0.45\linewidth}
        \centering
        \caption{Training.}
        \includegraphics[width=\linewidth]{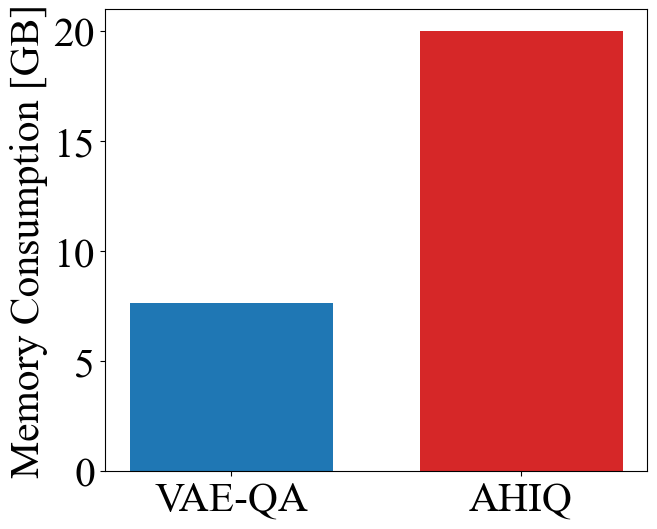}
        \label{fig:train_mem_consumption}
    \end{subfigure}
    \hfill
    \begin{subfigure}[b]{0.47\linewidth}
        \centering
        \caption{Inference.}
        \includegraphics[width=\linewidth]{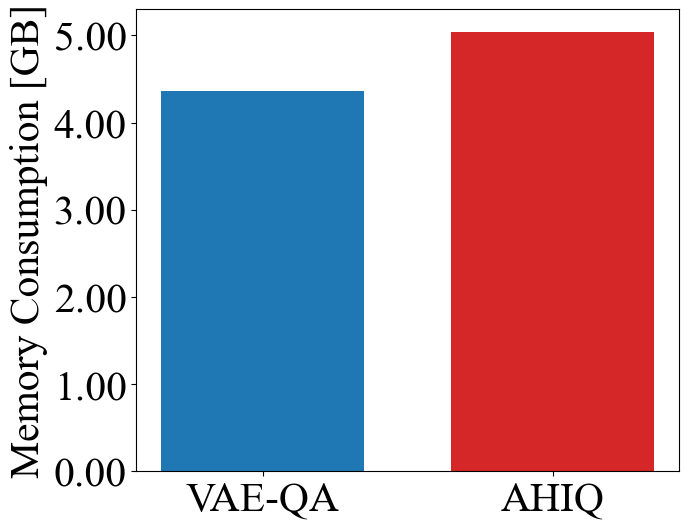}
        \label{fig:inference_mem_consumption}
    \end{subfigure}
    \caption{Memory footprint of the model.}
    \label{fig:mem_consumption}
    \vskip -0.2in
\end{figure}

In this section, we compare the memory footprint and runtime of our method with the current state-of-the-art method AHIQ~\cite{lao2022attentions}. We report the number of parameters and model size in \cref{tab:mem_consumption}. We also provide the model's memory consumption in \cref{fig:mem_consumption}. Finally, we report the model's runtime in \cref{tab:runtime}. This shows that our method has a smaller memory footprint and faster inference time than the current state-of-the-art method.

\begin{table}[H]
    \caption{Inference time per image.}
    \label{tab:runtime}
    \vskip 0.15in \centering\small\sc
    \begin{tabular}{@{}lc@{}}
        \toprule
        \textbf{Method} & \textbf{Runtime [ms]} \\
        \midrule
        AHIQ~\cite{lao2022attentions} & 35.16 \\
        \ourmethod{} (ours)           & 26.56 \\
        \midrule
        \textbf{Diff [\%]}            & 32.37 \\
        \bottomrule
    \end{tabular}
    \vskip -0.15in
\end{table}

\section{Conclusion}
\label{sec:conclusion}

We described \ourmethod{} a deep architecture for image quality assessment that is based on deep features learned using a generative (auto-encoding) task. The intuition is that generative latent representations are designed to capture the fine details of the image in contrast with discriminative approaches which preserve information about class labels. We find that \ourmethod{} consistently improves the accuracy of predicted quality compared with previous methods when tested on images from different datasets providing a new SoTA in this task. It also achieves a small improvement on average on same-dataset generalization. \ourmethod{} also has a smaller memory footprint and faster inference time than current SoTA.

These results suggest that current generative models can be easily leveraged for quality assessment. We expect that this approach can be generalized to other applications like video quality assessment.

\clearpage
\bibliographystyle{icml2024}
\bibliography{main}

\clearpage
\appendix

\section{Additional Results}
\label{sec:additional_results}

\subsection{Correlation by Distortion Type}
\label{sec:additional_results_correlation_by_distortion_type}

\begin{figure*}[ht]
    \vskip 0.2in \centering
    \includegraphics[width=1\textwidth]{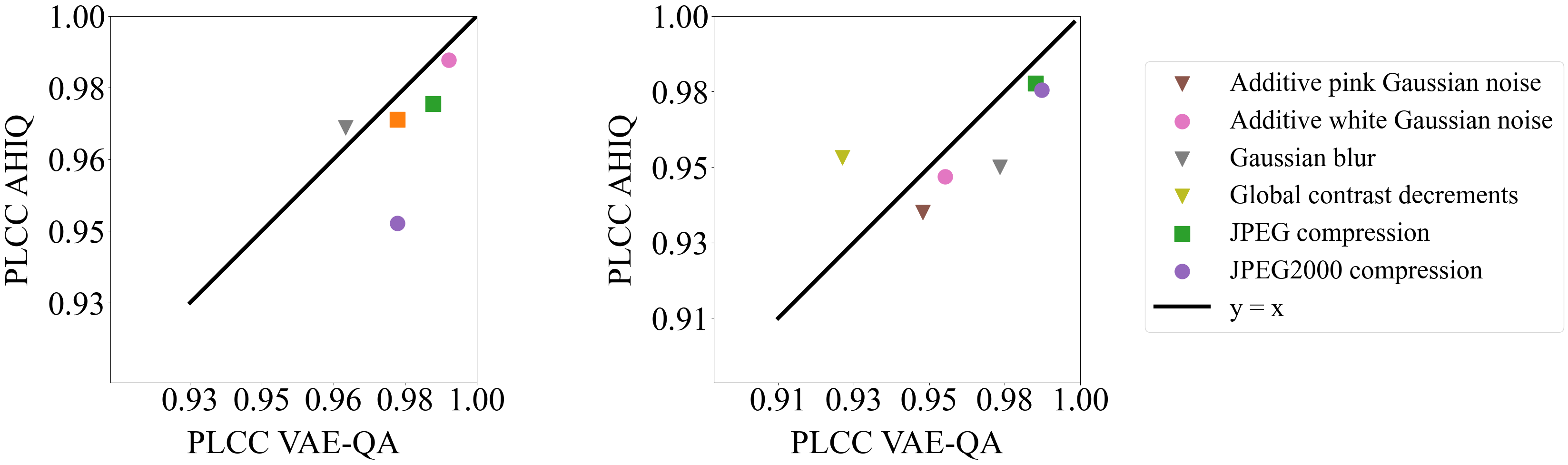}
    \caption{Quality prediction by distortion type on LIVE and CSIQ datasets. The figure compares PLCC obtained with our \ourmethod{} and AHIQ.}
    \label{fig:combined_distortion_types}
    \vskip -0.2in
\end{figure*}

To obtain more insight into the performance of \ourmethod{}, we measured the quality of quality prediction for different types of distortion. \cref{fig:combined_distortion_types} shows the performance of our method and AHIQ on the LIVE and CSIQ datasets. These datasets contain simpler distortions compared to TID2013, and \ourmethod{} reaches comparable performance to SoTA.

\subsection{Variability}
\label{sec:results_variability}

Tables \ref{tab:td_eval_std},\ref{tab:cd_eval_kadid10K_std}, and \ref{tab:cd_eval_tid2013_std} provide the standard deviation of our method across three seeds.

\begin{table*}[ht!]
    \caption{\textbf{Within-dataset evaluation - standard deviation.} Performance scores on three standard IQA datasets: LIVE, CSIQ, and TID2013. All scores are averages across three seeds. Scores for~\cite{lao2022attentions} were reproduced using authors code and are marked by (*).}
    \label{tab:td_eval_std}
    \vskip 0.15in \centering\small\sc
    \begin{tabular}{@{}lcc|cc|cc|cc@{}}
        \toprule
        \multicolumn{1}{c}{\textbf{\makecell[l]{Method}}} & \multicolumn{2}{c}{\textbf{ LIVE}} & \multicolumn{2}{c}{\textbf{ CSIQ}} & \multicolumn{2}{c}{\textbf{TID2013}} & \multicolumn{2}{c}{\textbf{ Overall}} \\
        \cmidrule(lr){2-3} \cmidrule(lr){4-5} \cmidrule(lr){6-7} \cmidrule(lr){8-9}
        & \makecell[c]{PLCC} & \makecell[c]{SRCC} & \makecell[c]{PLCC} & \makecell[c]{SRCC} & \makecell[c]{PLCC} & \makecell[c]{SRCC} & \makecell[c]{PLCC} & \makecell[c]{SRCC} \\
        \midrule
        AHIQ~\cite{lao2022attentions} (*) & \textbf{.988$\pm$.002} & \textbf{.985$\pm$.002} & .971$\pm$.008 & \textbf{.968$\pm$.008} & .927$\pm$.036 & .922$\pm$.045 & .962 & .958 \\
        \ourmethod{} (ours) & .984$\pm$.005 & .981$\pm$.004 & \textbf{.974$\pm$.009} & \textbf{.968$\pm$.009} & \textbf{.961$\pm$.016} & \textbf{.958$\pm$.017} & \textbf{.973} & \textbf{.969} \\
        \bottomrule
    \end{tabular}
    \vskip -0.15in
\end{table*}
\begin{table*}[ht!]
    \caption{\textbf{Cross dataset evaluation - standard deviation}. Trained on KADID-10k dataset and tested on LIVE, CSIQ, and TID2013 datasets. All scores are averages across three seeds. Scores for~\cite{lao2022attentions} were reproduced using authors code and are marked by (*).}
    \label{tab:cd_eval_kadid10K_std}
    \vskip 0.15in \centering\small\sc
    \begin{tabular}{@{}lcc|cc|cc|cc@{}}
        \toprule
        \multicolumn{1}{c}{\textbf{\makecell[l]{Method}}} & \multicolumn{2}{c}{\textbf{{KADID} $\rightarrow$ LIVE}} & \multicolumn{2}{c}{\textbf{KADID$ \rightarrow$ CSIQ}} & \multicolumn{2}{c}{\textbf{KADID$ \rightarrow$ TID2013}} & \multicolumn{2}{c}{\textbf{Overall}} \\
        \cmidrule(lr){2-3} \cmidrule(lr){4-5} \cmidrule(lr){6-7} \cmidrule(lr){8-9}
        &\makecell[c]{PLCC}  &\makecell[c]{SRCC}  &\makecell[c]{PLCC}   & \makecell[c]{SRCC} & \makecell[c]{PLCC} & \makecell[c]{SRCC} & \makecell[c]{PLCC} & \makecell[c]{SRCC} \\
        \midrule
        AHIQ~\cite{lao2022attentions} (*) & \textbf{.956$\pm$.004} & .954$\pm$.007 & .945$\pm$.005 & .940$\pm$.005 & .891$\pm$.003 & .888$\pm$.003 & .931 & .927 \\
        \textbf{\ourmethod{} (ours)} & .945$\pm$.001 & \textbf{.967$\pm$.001} & \textbf{.952$\pm$.001} & \textbf{.960$\pm$.001} & \textbf{.905$\pm$.003} & \textbf{.900$\pm$.003} & \textbf{.934} & \textbf{.942} \\
        \bottomrule
    \end{tabular}
    \vskip -0.15in
\end{table*}
\begin{table*}[ht!]
    \caption{\textbf{Cross dataset evaluation - standard deviation.} Trained on TID2013 dataset and tested on LIVE and CSIQ datasets. "CENTER" refers to using a single center crop. All scores are averages across three seeds.}
    \label{tab:cd_eval_tid2013_std}
    \vskip 0.15in \centering\small\sc
    \begin{tabular}{@{}lcc|cc|cc@{}}
        \toprule
        \multicolumn{1}{c}{\textbf{\makecell[l]{Method}}} & \multicolumn{2}{c}{\textbf{{TID2013} $\rightarrow$ LIVE}} & \multicolumn{2}{c}{\textbf{{TID2013} $\rightarrow$ CSIQ}} & \multicolumn{2}{c}{\textbf{Overall}} \\
        \cmidrule(lr){2-3} \cmidrule(lr){4-5} \cmidrule(lr){6-7}
        &\makecell[c]{PLCC}  &\makecell[c]{SRCC}  &\makecell[c]{PLCC}   & \makecell[c]{SRCC} & \makecell[c]{PLCC} & \makecell[c]{SRCC} \\
        \midrule
        VAE-QA (center) & .923$\pm$.001 & .942$\pm$.000 & .938$\pm$.002 & .935$\pm$.003 & .931 & .939 \\
        VAE-QA (ours) & \textbf{.930$\pm$.000} & \textbf{.948$\pm$.000} & \textbf{.943$\pm$.001} & \textbf{.941$\pm$.002} & \textbf{.937} & \textbf{.945} \\
        \bottomrule
    \end{tabular}
    \vskip -0.15in
\end{table*}

\section{Limitations}
\label{sec:limitations}

\subsection{Dataset Bias}
\label{sec:dataset_bias}

KADID-10k~\cite{lin2019kadid} was used for cross-dataset generalization, which is a large dataset with a range of distortions. However, the dataset does not necessarily represent the entire distortion distribution in reality. Furthermore, all datasets have a limited number of unique images, which makes generalization to real-world images difficult. Therefore, our method may suffer from some generalization errors when applied to real-world scenarios.

\subsection{Image Type}
\label{sec:image_type}

Our method uses the VAE from~\cite{rombach2021highresolution} to extract features from images. The VAE is trained on a OpenImages~\cite{Kuznetsova_2020} dataset, which contains natural images. Therefore, our method may not work well on images that are significantly different from natural ones, such as medical images, satellite images, or images of a particular domain.

\section{Implementation Details}
\label{sec:implementation_details}

This section describes in detail the components of our method.

\subsection{Feature Extraction Module}
\label{sec:implementation_details_feature_extraction_module}

We use feature maps from the VAE with the following dimensions: 
$f_{img} \in \mathbb{R}^{c_{0} \times r_{0} \times r_{0}}$, $f_{enc1} \in \mathbb{R}^{c_{1} \times r_{0} \times r_{0}}$, $f_{enc4} \in \mathbb{R}^{c_{1} \times r_{1} \times r_{1}}$, $f_{enc7} \in \mathbb{R}^{c_{2} \times r_{2} \times r_{2}}$, $f_{enc12} \in \mathbb{R}^{c_{3} \times r_{3} \times r_{3}}$, $f_{enc15} \in \mathbb{R}^{c_{3} \times r_{3} \times r_{3}}$, and $f_{enc19} \in \mathbb{R}^{c_{4} \times r_{3} \times r_{3}}$.

Where $c_{i}$ is the number of channels and $r_{i}$ is the spatial resolution of the feature map. Channel dimensions are $c_{i} = \{3, 128, 256, 512, 8\}$ and spatial resolutions are $r_{i} = \{256, 128, 64, 32\}$. The feature fusion module uses these feature maps as input.

\subsection{Feature Fusion Module}
\label{sec:implementation_details_feature_fusion_module}

The feature-fusion module combines features extracted from the VAE into a single unified representation. First, a difference feature map is created by subtracting the reference feature map from the distorted feature map. Then, those feature maps are passed to the fusion within VAE layer component.

\subsubsection*{Fusion within VAE layer:} 
\begin{algorithm}[H]
   \caption{$FusionWithin~VAE~Layer_{i}$}
   \label{alg:fusion_within_vae_layer}
\begin{algorithmic}
   \STATE {\bfseries Input:} feature map $F_{i}$
   \STATE {\bfseries Output:} fused feature map $F_{i}^{fused}$
   \STATE $h = Conv2d_{i}(F_{i})$
   \STATE $h = ReLU(h)$
   \STATE $h = Dropout2d(h)$
   \STATE $h = GroupNorm_{i}(h)$
   \STATE $F_{i}^{fused} = AdaptiveAvgPool2d(h)$
\end{algorithmic}
\end{algorithm}
The input to the fusion across VAE layers component is the concatenated feature maps:
\begin{align}
    F_{fused}^{ref} = \text{Concat}(\text{FusionWithinVAELayer}_{i}(F_{i}^{ref})), \\
    F_{fused}^{diff} = \text{Concat}(\text{FusionWithinVAELayer}_{i}(F_{i}^{diff})),
\end{align}
where ${F_{fused}^{ref}, F_{fused}^{diff}} \in \mathbb{R}^{\text{L*128} \times 16 \times 16}$, where L is the number of layers from the VAE plus the input image.

\subsubsection*{Fusion across VAE layers:}
\begin{algorithm}[H]
   \caption{$FusionAcrossVAELayers$}
   \label{alg:fusion_across_vae_layers}
\begin{algorithmic}
   \STATE {\bfseries Input:} feature map $F_{fused}$
   \STATE {\bfseries Output:} fused feature map $F_{fused}^{across}$
   \STATE $h = Conv2d(F_{fused})$
   \STATE $h = ReLU(h)$
   \STATE $F_{fused}^{across} = GroupNorm(h)$
\end{algorithmic}
\end{algorithm}

The input to the CNN flatten block is the concatenated feature map:
\begin{align}
    F_{fused}^{across} = \text{Concat}(&\text{FusionAcrossVAELayers}(F_{fused}^{ref}), \nonumber \\
    &\text{{FusionAcrossVAELayers}}(F_{fused}^{diff})),
\end{align}
Where $F_{fused}^{across} \in \mathbb{R}^{\text{1024*2} \times 16 \times 16}$.

\subsubsection*{CNN flatten block:} 
\begin{algorithm}[H]
   \caption{$FlattenBlock$}
   \label{alg:flatten_block}
\begin{algorithmic}
   \STATE {\bfseries Input:} feature map $F_{fused}^{across}$
   \STATE {\bfseries Output:} Compressed representation $C_{rep}$
   \STATE $h = Conv2d(F_{fused}^{across})$
   \STATE $h = ReLU(h)$
   \STATE $h = AdaptiveAvgPool2d(h)$
   \STATE $C_{rep} = Flatten(h)$
\end{algorithmic}
\end{algorithm}
The output of the feature fusion module is the compressed representation $C_{rep} \in \mathbb{R}^{1024*2*2}$.

\subsection{Quality Prediction Module}
\label{sec:implementation_details_quality_prediction_module}

\begin{algorithm}[H]
   \caption{$MLP$}
   \label{alg:mlp}
\begin{algorithmic}
   \STATE {\bfseries Input:} Compressed representation $C_{rep}$
   \STATE {\bfseries Output:} Quality prediction $q \in \mathbb{R}$
   \STATE $h = Linear(C_{rep})$
   \STATE $h = GroupNorm(h)$
   \STATE $h = ReLU(h)$
   \STATE $h = Dropout(h)$
   \STATE $h = Linear(h)$
   \STATE $h = ReLU(h)$
   \STATE $q = Linear(h)$
\end{algorithmic}
\end{algorithm}

\subsection{Mapping DMOS to MOS}
\label{sec:mapping_mos_to_dmos}

 Image quality can be assessed using \textit{Mean Opinion Score} (MOS) and \textit{Difference MOS} (DMOS) based on human opinions. For FR-IQA, MOS is the average score a group of people give after seeing an image compared to its original version. The DMOS is the difference between the raw quality scores of the reference and distorted images. The DMOS is calculated by subtracting the MOS of the reference image from the MOS of the distorted image~\cite{mohammadi2014subjective}. DMOS measures the impact of a particular distortion on image quality, while MOS measures overall quality.
 
For datasets that provide DMOS values, we transform them to MOS using the following formula:
\begin{align}
    \text{{MOS}} = \text{{max}}(\text{{DMOS}}) - \text{{DMOS}}
\end{align}

\section{The Effect of Saliency}

It has been shown that predicting perceptual quality can be improved by integrating saliency maps, which capture how much attention people pay to various areas in a given image~\cite{seo2020novel}.

Adding salience is likely to improve other image quality assessment than the ones tested in~\cite{seo2020novel}, because it helps assessment methods focus on the most important parts of an image. Unfortunately, we were unable to obtain the code from the authors and could not make a direct comparison.

\end{document}